\documentclass[conference]{IEEEtran}
\IEEEoverridecommandlockouts
\usepackage{cite}
\usepackage{amsmath,amssymb,amsfonts}
\usepackage{algorithmic}
\usepackage{graphicx}
\usepackage{textcomp}
\usepackage{xcolor}
\usepackage{xcolor}
\usepackage{soul}
\usepackage{url}
\usepackage{booktabs}

\def\BibTeX{{\rm B\kern-.05em{\sc i\kern-.025em b}\kern-.08em
    T\kern-.1667em\lower.7ex\hbox{E}\kern-.125emX}}
\begin{document}

\title{Automated computed tomography and magnetic resonance imaging  segmentation using deep learning: a beginner's guide\\
\thanks{We thank the following research support foundations: CAPES, CNPQ, and grant \#2019/21964-4, São Paulo Research Foundation (FAPESP).}
}

\author{\IEEEauthorblockN{Diedre Carmo\textsuperscript{\textsection}}
\IEEEauthorblockA{\textit{MICLab} \\
\textit{University of Campinas (UNICAMP)}\\
Campinas, Brazil \\
d211492@dac.unicamp.br
}
\and
\IEEEauthorblockN{Gustavo Pinheiro\textsuperscript{\textsection}}
\IEEEauthorblockA{\textit{MICLab} \\
\textit{University of Campinas (UNICAMP)}\\
Campinas, Brazil \\
g162793@dac.unicamp.br}
\and
\IEEEauthorblockN{Lívia Rodrigues\textsuperscript{\textsection}}
\IEEEauthorblockA{\textit{MICLab} \\
\textit{University of Campinas (UNICAMP)}\\
Campinas, Brazil \\
l180545@dac.unicamp.br}
\and
\IEEEauthorblockN{Thays Abreu}
\IEEEauthorblockA{\textit{MICLab} \\
\textit{University of Campinas (UNICAMP)}\\
Campinas, Brazil \\
thays@unicamp.br}
\and
\IEEEauthorblockN{Roberto Lotufo}
\IEEEauthorblockA{\textit{MICLab} \\
\textit{University of Campinas (UNICAMP)}\\
Campinas, Brazil \\
lotufo@unicamp.br}
\and
\IEEEauthorblockN{Letícia Rittner}
\IEEEauthorblockA{\textit{MICLab} \\
\textit{University of Campinas (UNICAMP)}\\
Campinas, Brazil \\
lrittner@unicamp.br}

}

\newcommand{\livia}[1]{\textcolor{pink}{{#1}}}
\newcommand{\diedre}[1]{\textcolor{green}{{#1}}}
\newcommand{\gustavo}[1]{\textcolor{blue}{{#1}}}
\newcommand{\thays}[1]{\textcolor{orange}{{#1}}}

\maketitle
\begingroup\renewcommand\thefootnote{\textsection}
\footnotetext{Equal contribution}
\endgroup
\begin{abstract}

Medical image segmentation is an increasingly popular area of research in medical imaging processing and analysis. However, many researchers who are new to the field struggle with basic concepts. This tutorial paper aims to provide an overview of the fundamental concepts of medical imaging, with a focus on Magnetic Resonance and Computerized Tomography. We will also discuss deep learning algorithms, tools, and frameworks used for segmentation tasks, and suggest best practices for method development and image analysis. Our tutorial includes sample tasks using public data, and accompanying code is available on GitHub (https://github.com/MICLab-Unicamp/Medical-Imaging-Tutorial). By sharing our insights gained from years of experience in the field and learning from relevant literature, we hope to assist researchers in overcoming the initial challenges they may encounter in this exciting and important area of research.
\end{abstract}

\section{Introduction}

It is not a mystery that medical images are a great tool in medicine. Besides being non-invasive, they are useful for diagnosing, evaluating, and predicting diseases. Also, many physicians use medical images for research purposes. 
The history of medical imaging begins in 1895, with the discovery of X-rays by Wilhelm Rontgen. It did not take long, the technique was being used by physicists to analyze medical issues. However, only in the later 1960's Hounsfield, an EMI Limited researcher, started to study x-rays in a 3D form. In 1972, Hounsfield and Dr. James Ambrose could diagnose a tumor using Computerized Tomography (CT) due to different tissue contrasts. By the same time, in 1973, Paul Lauterbur demonstrated that Nuclear Magnetic Resonance (NMR) could be used to create an image \cite{lauterbur1973image}. Yet, only in 1977 the first human image using MR was acquired, taking 5 hours for the acquisition\cite{edelman2014history}.

With the popularization of imaging methods, research centers are dealing with an increasing amount of data that are both time and monetary costly to analyze manually. Using computational methods, engineers and computer science professionals can help physicians diminish these costs.
In the early days, these computational methods were mainly for image enhancement, based on classical imaging algorithms such as morphology and filtering.
More complex tasks, such as pattern recognition through machine learning algorithms, became possible as the area evolved.
Although the first work on artificial intelligence (AI) began in the 1950s, AI only started to be applied to medical issues in the 1970s \cite{kaul2020history}.

By the decade of 1990, there were several segmentation methods applied to MR images, and it was common to find methods based on the image characteristics, such as region-based algorithms. After a while, machine learning (ML) methods started to appear, and it became more usual to extract features from the images and use them on algorithms such as Support Vector Machines or Decision Trees. Finally, deep learning (DL) has become central over the last few years. Unlike the previous methods applied to medical images, DL can compute features from the raw input data without requiring manual feature extraction. The feature extraction step is done by the first layers of the architecture, and it is not necessarily interpretable by humans since they are abstract representations of the input image. However, DL algorithms require a large amount of data, which is a challenge in the medical field. 


DL algorithms are widely applied to various medical image problems, such as regression, classification, and segmentation. In classification problems, DL classifies a sample into one of the N possible labels. Many applications in the literature focused on helping medical research, such as brain tumor classification~\cite{aurna2022classification,deepak2019brain} and covid $ \times$ non-covid affected lungs method~\cite{carmo2021rapidly}. 
In regression problems, the model is trained to predict a continuous value. Bounding box prediction for structures detection~\cite{gauriau2015multi,noothout2020deep} and MRI reconstruction from K-space images~\cite{beauferris2022multi} are examples of regression used on medical images. Finally, segmentation methods contour the border of a specific structure or area to be studied. Physicians may use the segmentation to study brain lesions~\cite{pinheiro2018v}, lung findings~\cite{carmo2021multitasking}, and subtle changes in specific structures. Several DL networks have been developed for this purpose \cite{roy2019quicknat, ronneberger2015u} and are applied to segment different structures of the brain~\cite{carmo2021hippocampus,rodrigues2020hypothalamus} and body~\cite{rahman2020automatic, tang2020two}. 

This paper will focus on Magnetic Resonance (MR) and Computed Tomography (CT) segmentation using DL algorithms. First, we will give a brief overview on the data, its acquisition, and intrinsic characteristics. Then, we will shortly introduce DL for image analyses and the computational environment required. Finally, we will describe the usual workflow and the most common statistical analysis, and give some recommendations and useful tips.


\section{Medical images}

A medical image can be understood as any image that represents aspects of the biological tissue. Medical images can be classified by the technique that is used for acquisition, also called modalities, such as ultrasound, magnetic resonance, and X-ray computed tomography. They can also be classified by their dimension: planar (or 2D), volumetric (or 3D), time series (4D), or by their range of values: single-channel scalar; multi-channel (e.g. dermatoscopic image) scalar; tensorial (e.g.: diffusion tensor imaging).

Among the great variety of medical image types, the vast majority is composed of scalar measures.
The images are made of a collection of voxels (volume elements) representing a single scalar value. Thus, the images are a scalar field in a $\mathbb{Z}^3$ space. 



\begin{figure}[ht]
    \centering
    \includegraphics[width=1.0\columnwidth]{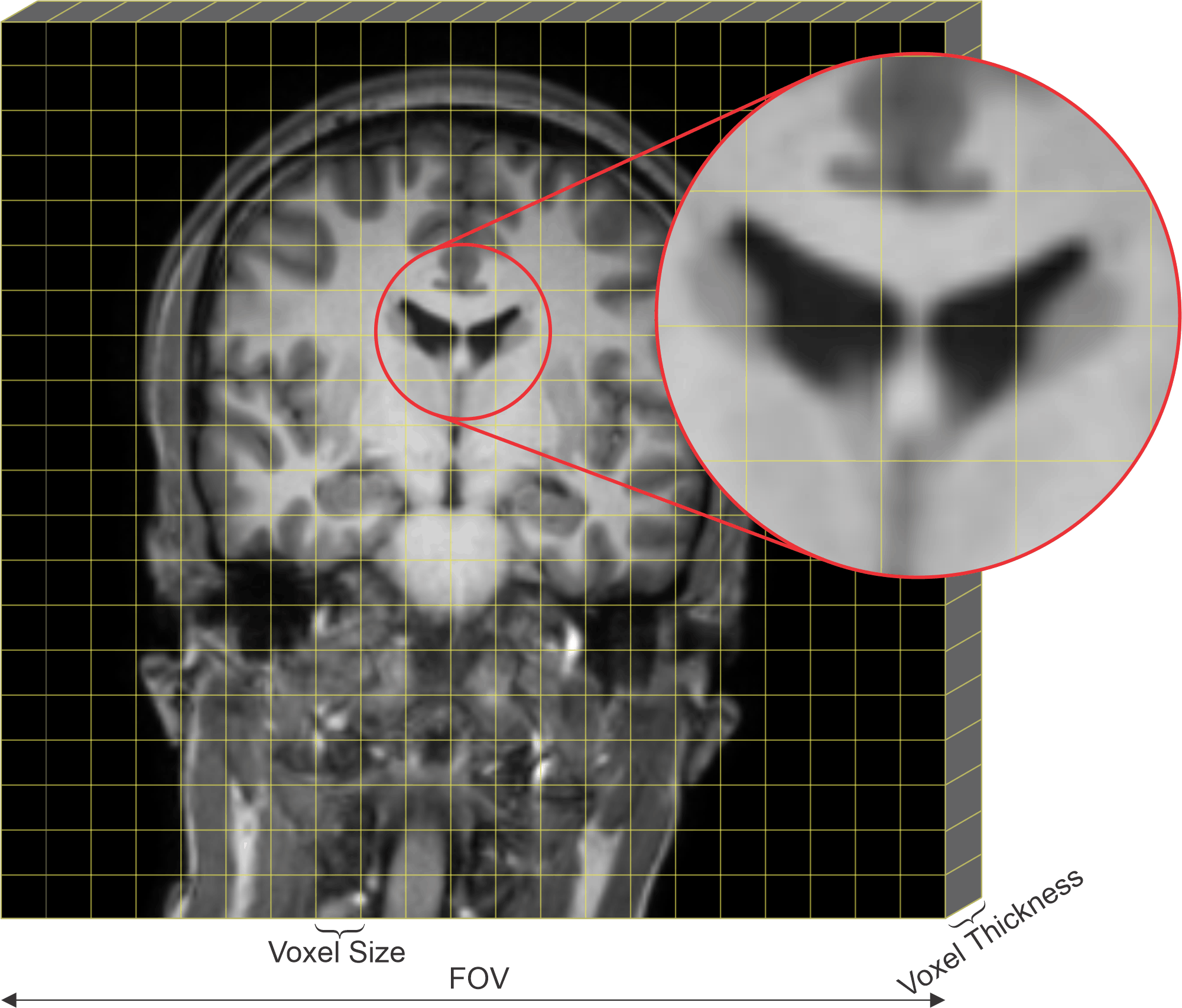}
    \caption{\label{fig:FOV}Parameters of a 3D image: Fiel of View (FOV), resolution, and voxel size.}
\end{figure}

As for any finite discrete scalar field, or array, the images are defined by geometrical parameters that are directly related to the image quality. Field of View (FOV), spatial resolution, Voxel Size (Fig.~\ref{fig:FOV}), and radial resolution are some of the most relevant parameters. The FOV is the size of the image in real-world dimensions, the spatial resolution is the number of voxels in each image dimension, the voxel size is the measurement of the voxel dimensions, and the radial resolution is the number of possible scalar values that the voxel can assume. This scalar value is the representative of the tissue for each voxel.

\begin{figure}[ht]
    \centering
    \includegraphics[width=1.0\columnwidth]{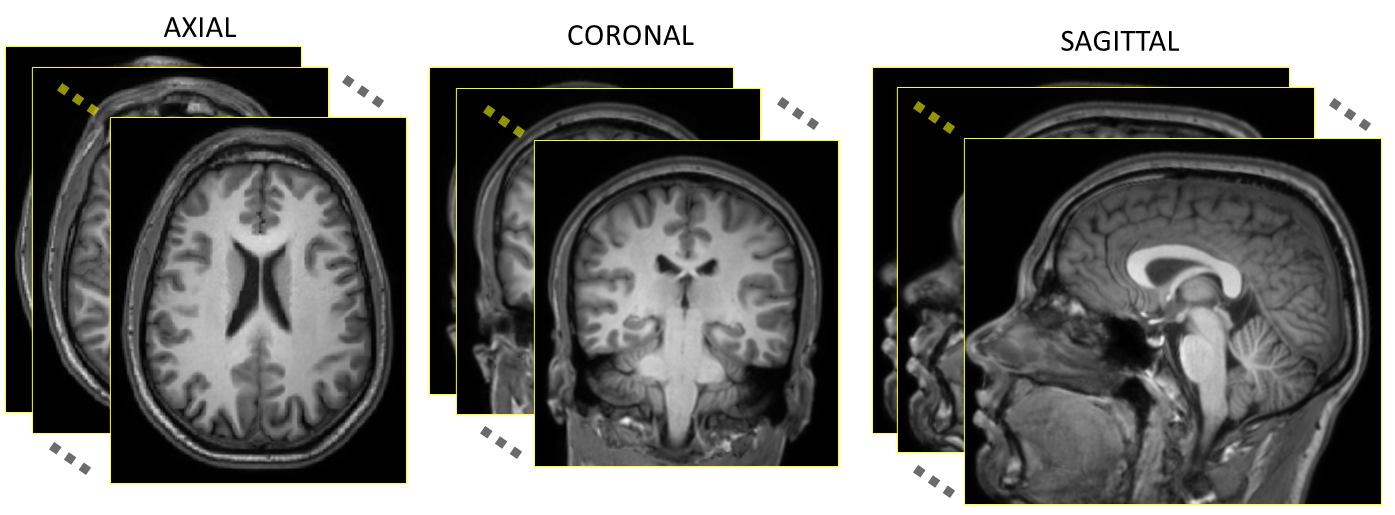}
    \caption{\label{fig:views}Slice view orientation in 3D medical images: Axial, Coronal, and Sagittal}
\end{figure}

As medical images are usually acquired in 2D slices, the 3D image can be seen as a stack of slices, and a 3D image gives us the freedom to look at the image from three views~(Fig.~\ref{fig:views}): axial (or transversal), coronal (or frontal), and sagittal (or longitudinal). 

The voxel is a volumetric element expressed in a single scalar value, and it represents an average of what is inside the defined space. Consequently, the voxel value suffers from the partial volume, which happens when more than one biological tissue is represented in a single voxel. This effect is minimized by reducing the voxel size and increasing the resolution for the same FOV. The partial volume can affect both manual and automatic segmentation methods since the fuzziness of the values could confuse the algorithms.

Depending on the application, the image acquisition parameters could vary considerably. For example, in the research field, the images using research-grade parameters tend to have higher resolution, better contrast, and less noise at the cost of longer acquisition time or better quality equipment. On the other hand, for clinical purposes, where resources such as time are scarce, the images acquired in clinical settings usually have a less spatial resolution (sometimes even skipping slices) and lower quality in general.

Among all 3D medical imaging types, CT and MR imaging are two of the most popular imaging-based diagnostic modalities used in different clinical conditions for diagnosis, follow-up, image-guided procedures, and medical research. Studying these images allows the analysis and segmentation of different body structures. Annotations can be used for different purposes such as volume measurements, statistics of a population in medical research, localization of abnormal tissue according to an underlying pathological process, and disease of the patient~\cite{buzug2011computed}. Since manual analysis is very time-consuming and poorly reproducible, there is an interest in automated processing of these images~\cite{suzuki2017overview}.

\subsection {X-ray CT} 

\begin{figure}[ht]
    \centering
    \includegraphics[width=\columnwidth]{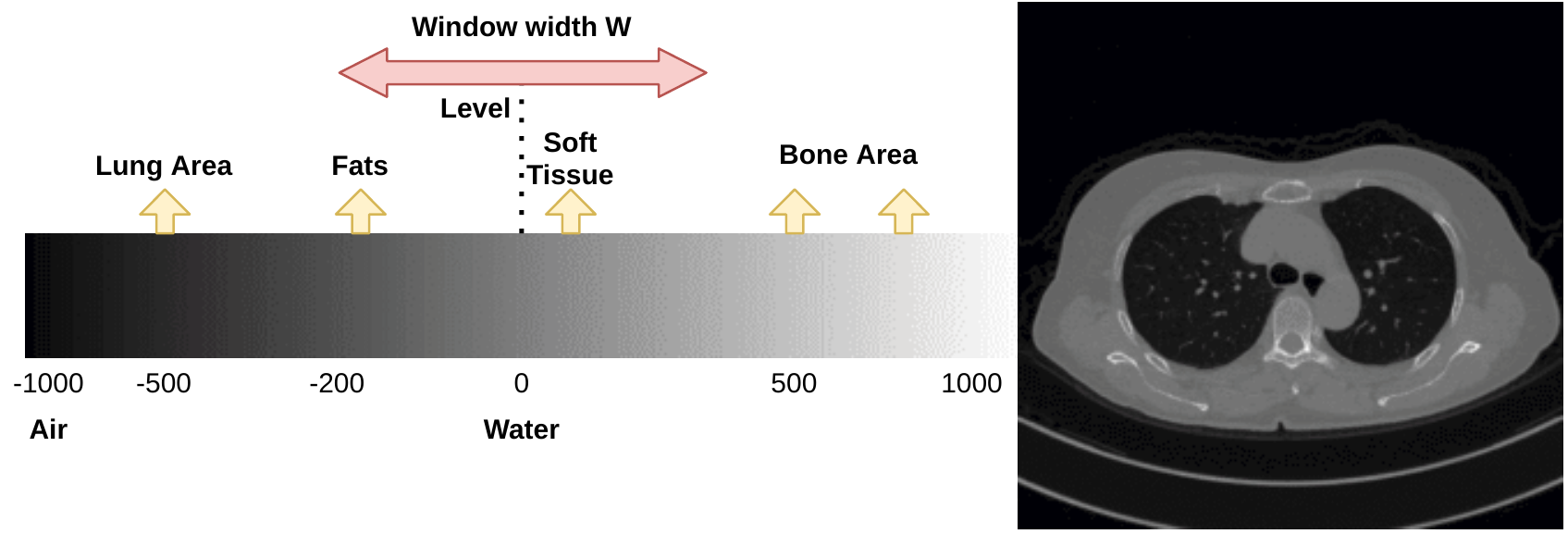}
    \caption{\label{fig:hu}Hounsfield Unit scale, mapping values to the represented tissue, and an axial slice of a CT scan showing the inside of the lung.}
\end{figure}

In the X-ray based CT modality, the image is reconstructed from various X-ray acquisitions around the patient. Several parameters guide the acquisition and reconstruction of signals recorded by the CT scanner into the final image and can be found on the image's header. These parameters include, among others: the spacing between body slices, which can range from less than 1 mm to more than 1 cm; slice resolution, which is commonly very high with a pixel representing only 0.5mm²; reconstruction kernel or filter, which controls the frequencies present in the image and can generate from very smooth to very noisy images; and many others. One of the most important parameters to be considered when processing CT images is the Hounsfield Units (HU) window. HU directly maps to specific tissues, air, and water (Fig~\ref{fig:hu}). Other variants of CT acquisition, such as PET-CT are out of the scope of this manuscript.

Since HU values go beyond the traditional 8-bit (256 values) representation of gray images in a monitor, applying a window (clipping) to HU values will improve the contrast and visualization of specific tissues. In addition, the values contained in the digital reconstruction may have been rescaled to a different value. This rescaling is represented by a linear mapping $ax + b$ where $a$ is the rescale slope and $b$ is the rescale intercept. These are commonly applied to remove negative numbers from the intensities, allowing unsigned storage. Section~\ref{sec:preprocess} will review some recommendations to leverage these proprieties of CT images and avoid common mistakes when preprocessing and using CT images.

\subsection {MRI}

MR Imaging is a technique that creates images by the interference between a high-intensity magnetic field, radio frequency pulses, and the field generated by the spin of the protons in the tissues inside the MR scanner. The signal strength measured by the scanner coils is responsible for generating the contrast for each voxel of the three-dimensional image.

MR images play a key role in the clinical environment as it is proven to be a fast, safe (non-ionizing radiation), and non-intrusive way to look inside the body. To support different types of exams, there are several MR imaging pulse sequences, each one presenting a different image contrast, exploiting specific characteristics of biological tissues. 
Due to a qualitative similarity to an anatomical slice, the sequence known as T1-weighted is one of the most popular. It is also one of the fastest-acquired sequences and relatively simple to analyze.

\begin{figure}[ht]
    \centering
    \includegraphics[width=1.0\columnwidth]{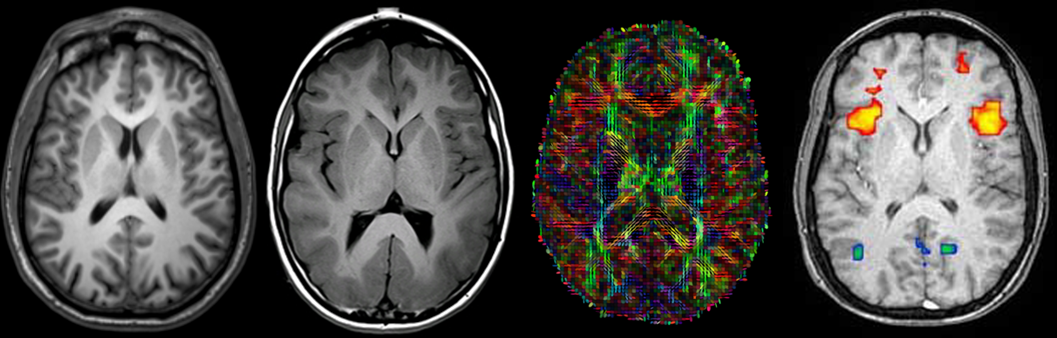}
    \caption{\label{fig:sequences}Examples of MRI images from different acquisition sequences: respectively T1-weighted, FLAIR, Diffusion Tensor Imaging, and Functional MRI. Images from different subjects and non-correspondent slices.}
\end{figure}

Besides T1-weighted images, many other MR image sequences (Fig.~\ref{fig:sequences}) are used for various applications, including structural and functional. For example, FLAIR~\cite{kilsdonk2014improved} is used for investigating brain lesions, Diffusion sequences are used to measure the water diffusion in the tissue and infer about the fiber organization~\cite{basser1995diffusion}, Functional MR imaging \cite{huettel2004functional} is used to measure brain activity, and spectroscopy \cite{harris1986nuclear} to measure metabolites.

\subsection{Medical Imaging Processing}
  
The field of medical imaging processing has considerably changed over the last few years. At first,  methods were based on image characteristics and often mimic the process done by human experts, such as region-based algorithms. After a while, machine learning (ML) methods started appearing, and extracting features from the images and using them in the algorithms became more usual. Finally, over the last few years, deep learning (DL) has taken the central place~\cite{carmo2022systematic}.



Complementary to previous machine learning methods, Deep Learning can learn how to generate the features automatically. A Deep Learning model can perform well using only raw data as input, without the necessity of a previously manually engineered feature extraction section, even requiring less technical knowledge about the problem in question to achieve baseline results~\cite{lecun2015deep}.

In the medical imaging context, convolutional kernel parameters are the weights adjusted in training. The application of these convolutional kernels is invariant to translation, which allows learning from different positions throughout the image. Optimizing the kernel weights takes place by minimizing a loss function~\cite{sudre2017generalised} expressing the model performance, through the application of optimization methods using variations of stochastic gradient descent~\cite{DBLP:journals/corr/KingmaB14}. In addition, these convolutions are usually supported by normalization layers~\cite{DBLP:journals/corr/IoffeS15} and nonlinear activations~\cite{xu2015empirical}. Normalization and non-linearity were some of the main breakthroughs that allowed deep learning networks to thrive in many applications, significantly accelerating the convergence of training. The Deep Learning architectures that use these techniques are called Convolutional Neural Networks (CNN). Note that CNNs are not the only way to use deep learning for imaging applications, with Tranformers~\cite{shamshad2022transformers} also being a current competitive architecture paradigm in natural and medical imaging processing. 

There is an infinite amount of possible combinations of convolutional and other types of layers to define a CNN architecture. However, some combinations, or architectures, are already defined and well-known for certain applications. For example, U-Net~\cite{ronneberger2015u} is one of the most used architectures in the field of medical imaging for image segmentation due to its compact encoder-decoder design that propagates multi-resolution features from the encoder to the decoder, allowing for fewer parameters when compared to famous large natural image classification architectures such as ResNet~\cite{he2016deep}, EfficientNet~\cite{tan2019efficientnet}, and ConvNext~\cite{liu2022convnet}. The encoder acts as a feature extractor and the decoder as a reconstructor of the intended output. It is important to stress that those features automatically extracted by the encoder are not necessarily visually interpretable since they are abstract representations of the input image that maximize a given objective. From the seminal UNet paper, many variations of the encoder-decoder segmentation architecture have been proposed~\cite{chen2018encoder, abdollahi2020vnet}, and the encoder-decoder design continues to be the de-facto approach for supervised automated medical imaging segmentation.

\section{Environment Recommendations}
Software and hardware recommendations can be sensitive since different people and groups will have different experiences. In this Section, we will recommend what has worked for us in recent years, focusing on DL over medical imaging processing. However, these recommendations can be extrapolated using different processing methodologies. Firstly, regarding hardware, although having a good amount of RAM and a good CPU helps, we recommend distributing your budget, focusing on the best possible GPU with a decent amount of video memory. We also suggest paying attention to how much storage your work needs since some medical images can use a large amount of storage. If possible, Use a high-speed (SSD) storage for preprocessed data and low-speed storage for raw/original data.

From our experience, we recommend using Python and PyTorch as the programming language and DL framework, respectively. PyTorch~\footnote{https://pytorch.org/} follows an object-oriented programming approach and is currently being used by a large part of the DL community. A top-level framework that simplifies some of the ``engineering" code necessary to use PyTorch is PyTorch Lightning~\footnote{https://pytorchlightning.ai/}. The libraries you will use for processing will vary depending on the needs of your project. Some libraries commonly used for reading and dealing with medical imaging include NumPy, SimpleITK, Nibabel, and Pydicom. Both Windows and Ubuntu work well with Python in terms of operational systems. Jupyter Notebooks are a good tool for a more interactive programming approach, especially for prototyping and proof of concept. As a workflow, having separate Python scripts that are imported in a jupyter notebook or a main command line script is interesting for controlling input arguments. Finally, we recommend that you use logging frameworks for experiments and do not implement logging from scratch. Many useful tools are available to log experiment parameters and results, and they are essential for organizing your research and recalling experiments in the future. Some examples include Neptune.ai\footnote{https://neptune.ai/} and TensorBoard\footnote{https://www.tensorflow.org/tensorboard}.

Storage requirements are one of the main things to keep in mind when dealing with digital medical imaging. Unlike other imaging processing areas that deal with 2D images, medical imaging is frequently three-dimensional, with high resolution. This results in uncompressed storage of one volume, sometimes using hundreds of megabytes of space. Therefore, you need to reserve gigabytes or even terabytes of space to store both original and preprocessed copies of your data. We recommend that preprocessed data be stored in fast storage for faster processing, such as SSDs, and original data be stored in slower HDDs or even a separate computer. For deep learning training, using GPUs is mandatory, given that the training process consists of parallel multiplications and sums that can be performed optimally in GPUs. We recommend using a modern Nvidia GPU with CUDA support, due to it being the norm in the field currently, with most frameworks using CUDA. For RAM, We recommend a minimum of 16 GB of RAM, with more being beneficial especially for training 3D networks. Having more RAM is not essential but beneficial, increasing the amount of data that can be cached during training, and reducing the possibility of bottlenecks in parallel data loading. There is no need to have the most expensive CPU, but a top tier CPU will be of benefit for data processing and loading during training. Finally, it is very important to highlight that the GPU usage during training should be close to 100\%. If not, that means that data loading is a bottleneck, either due to slow readings from storage, slow/not enough RAM, or a slow CPU. Always pay attention to hardware cooling as modern Nvidia GPUs should stay below 83 degrees Celsius while training. Finally, try to use parallel data loading and optimize your code. Badly optimized data loading code can bring even the most expensive hardware to a halt in training.

\section{Workflow}

\begin{figure*}[t]
    \centering
    \includegraphics[width=\textwidth]{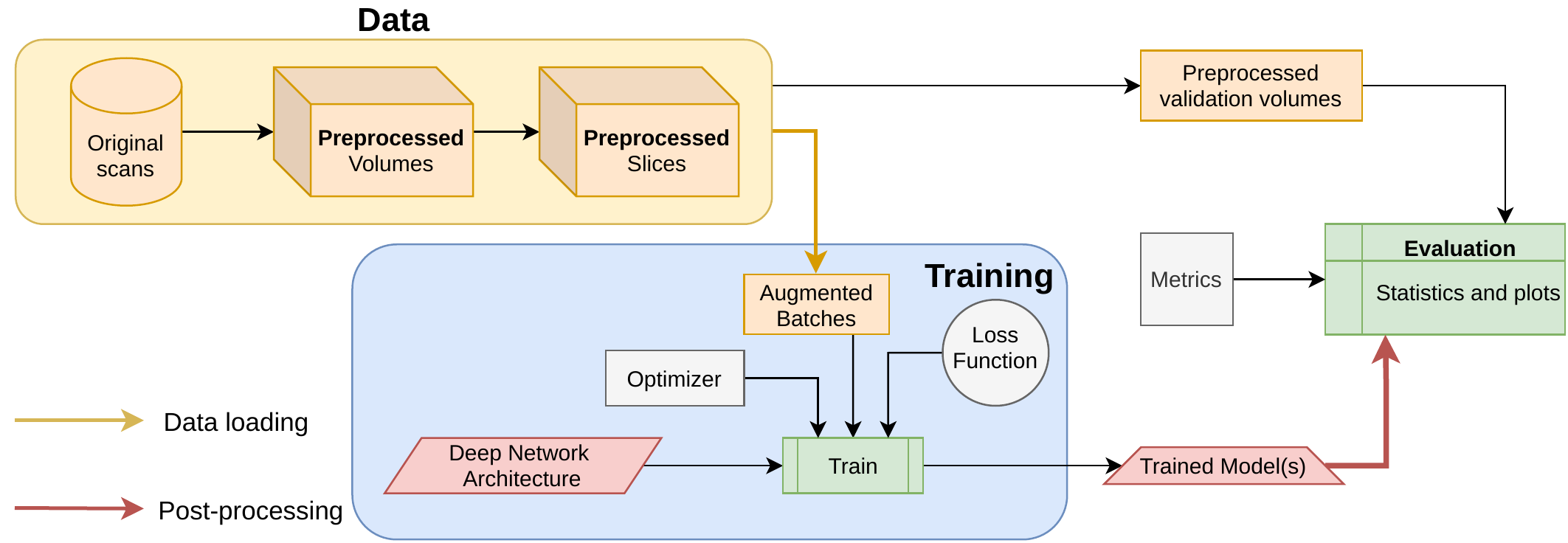}
    \caption{\label{fig:workflow}Illustration of a generic workflow for training DL models for medical imaging analysis, from data pre-processing to model training and evaluation.}
\end{figure*}

This section will go through recommendations and tips on all steps of the DL for medical imaging analysis workflow (Fig.~\ref{fig:workflow}). 

\subsection{Data}

\begin{table*}[ht]
\resizebox{\textwidth}{!}{
\begin{tabular}{llll}
\toprule
\textbf{Dataset}                       & \textbf{Annotation}                    & \textbf{Description}                                & \textbf{Number of images}                                         \\\midrule
\multicolumn{4}{c}{\textbf{CT}} \\\midrule

 LTRC \cite{karwoski2008processing}     & --                                     & Lung Tissue Research Consortium                     & 1200 patients                                                               \\
 LIDC-IDRI\cite{armato2011rd}                        & Nodule Detection                       & Lung Image Database Consortium Image Collection     & 1018 scans                                                                             \\
 LOLA11\cite{lola11}                           & Lobe masks (external)                  & LObe and Lung Analysis Challenge                    & 55 scans                                                                               \\
 LUNA16\cite{luna16}                           & Pulmonary nodules                      & Challenge for pulmonary nodule segmentation         & 888 scans                        \\
 IEE CCAP\cite{yan2020ccap}                         & --                                     & COVID-19 low dose scans                             & 154 scans                                                                                \\
 MSC\cite{msl}                       & Lung and COVID-19 findings             & COVID-19 patients scans                             & 100 scans                        \\
 MOSMED\cite{morozov2020mosmeddata}                           & COVID-19 findings                      & COVID-19 patients scans                             & 860 slices                                                                               \\
 CoronaCases \cite{jun2020covid}                      & Lung and COVID-19 findings             & COPD and COVID-19 patients scans                    & 20 scans                                                                                 \\
 MSD \cite{jun2020covid}    & Pulmonary nodules segmentation         & Challenge with multiple tasks including lung cancer & 2633 slices                  \\
 MEDGift\cite{kumar2018segmentation}             & ILD annotations                        & Multimedia dataset of ILD cases                     & 128 patients                                                                            \\
 Empire10\cite{murphy2011evaluation}                         & --                                     & Registration of thoracic CT data                    & 30 scan pairs                                                                            \\
 VESSEL12\cite{vessel12}                          & Lung and vessel                        & VESsel SEgmentation in the Lung Challenge           & 30 scans                                                                                 \\
 VISCERAL\cite{langs2012visceral}                          & Various modalities including lung Lung & Visual Concept Extraction Challenge in Radiology    & 80 volumes                                                                               \\
 DSB\cite{dsb}     & Cancer                                 & Data Science Bowl 2017 Kaggle Competition           & 2101 scans                                                                               \\
 Kaggle\cite{kaggle}            & Lung annotations                  & Kaggle lung segmentation challenge                  & 267 slices                                                                               \\
 NSCLC\cite{wang2016automatic}                    & Pulmonary nodule annotations           & Non-small cell lung cancer patients                 & 422 patients                                                                              \\
 EXACT09\cite{lo2012extraction}                          & Airway annotations                     & Extraction of Airways from CT                       & 40 scans                                                                                 \\
 ImageCLEF\cite{cid2015efficient}                    & Lung masks                             & Tuberculosis severity scoring challenge             & 335 scans                                                                                \\
 SARS-CoV-2\cite{soares2020explainable}                        & Lung                                   & COVID-19 patients scans                             & 2482 scans                                                                               \\
 TCIA\cite{tcia}            & Lung and COVID-19 findings             & COVID-19 patients scans                             & 753 patientes                                                                           \\\midrule
 \multicolumn{4}{c}{\textbf{MRI}} \\\midrule
 ADNI~\cite{jack2008alzheimer} & -- & Patient (Alzheimer and mild cognitive impairment)  and controls & 1821 scans \\
 OASIS~\cite{lamontagne2019oasis} & -- & T1-w, T2-w, FLAIR, ASL, SWI, time of flight, resting-state BOLD, and DTI & $>$2000 scans \\
 HCP~\cite{van2012human} & -- & Human Connectome Project: Structural and diffusion data & 1200 patients \\
 IXI~\cite{IXI} & --  & Information eXtraction from Images & 581 patients \\
 ABIDE~\cite{di2014autism} & -- & Autism Brain Imaging Data Exchange & 1112 patients \\
 Kirby21~\cite{landman2011multi} & -- & Test-retest dataset on 21 subjects & 42 scans \\ 
 HARP~\cite{frisoni2015eadc} & Hippocampus mask & EADC-ADNI Harmonized Protocol for manual hippocampal segmentation on magnetic resonance &  135 patients\\
 CC359~\cite{souza2018open} & Automated and manual Skull Stripping & Healthy subjects T1-w MRI & 359 patients \\
 BRATS~\cite{brats} & Brain tumor manual segmentation & The multimodal brain tumor image segmentation benchmark & 65 scans  \\
 MiLI~\cite{rodrigues2022benchmark} & Automated and Manual hypothalamus masks & MICLab-LNI Initiative: T1-w MRI & 452 patients \\
 Thalamus~\cite{} & Automated and Manual thalamus masks & T1-w and Diffusion MRI & 1063 patients \\
 LBPA40~\cite{shattuck2008construction} & 56 brain structure masks & 3D probabilistic atlas of human cortical structures & 40 patients \\
 NFBS~\cite{puccio2016preprocessed} & Manual skull stripping & Neurofeedback Skull-stripped: T1-w MRI & 125 scans \\
 IBSR~\cite{ibsr} & 43 structures of the brain & Internet Brain Segmentation Repository & 18 patients \\
 LGG~\cite{buda2019association} & Manual FLAIR abnormality segmentation masks & Brain MRI from patients with lower-grade gliomas & 110 patients \\
 
 IBSD\cite{sivaswamy2021sub} & Subcortical segmentation & Indian Brain Segmentation Dataset & 114 patients \\\bottomrule
\end{tabular}
}
\caption{\label{tab:data}Examples of public datasets of MRI or CT scans, some with included voxel-wise segmentation annotations.}
\end{table*}

Data curation is one of the most important steps of the training workflow. Here, we list five steps you need to consider before training your network.


\subsubsection{Understanding your problem and your data}

The first thing you need to do is to understand the data you are using. For instance, it will be hard to classify lung lesions in CT images if you cannot locate the lung or if you have not heard about Hounsfield units. The more you understand the problem, the easier it is to find computational solutions for it. For example, if you are trying to segment a small structure, you may need to use some specific loss function or pay attention to the depth of the network.
In this first step, you need to talk with physicians, radiologists, and specialists about the problem you are working on. You need to understand the primary goal of your project. Are you trying to decrease the processing time of existing methods, to improve metrics regardless of the processing time, or maybe, to develop a more generic application, even if it means a loss in performance? 
For instance, if your application is meant to work only in one imaging center, always using the same type of acquisition, you may train your network using only one dataset. However, if you intend to develop an application that can be used by different research centers, it is imperative to mix different data sources to increase generalization.

By understanding your problem early in development, you will be able to find the most effective data for your application and avoid wasting time re-training in the future.

\subsubsection{Labels}

Here, you need to define what will be your ground truth. Usually, semi-supervised and supervised applications will require trustful labels to deliver good predictions. For medical images, the gold standard is usually defined as a specialist annotation. Also, the ideal scenario is to have more than one specialist annotating your data to avoid bias.
However, the cost of manual labels is high in terms of both time and money. So, the first thing you need to do before starting a project is to analyze if you have data and annotation.

In terms of segmentation, if you have only a few labeled data, you may try some tricks:

\begin{itemize}
    \item \textit{Silver Standards}: Manual annotation is time-consuming, so you may create a dataset using established automated methods (Tab.~\ref{tab:data_lab}). To deal with different types of segmentation and reduce label noise, it is recommended to use label-cleaning strategies such as STAPLE and majority voting. Souza \textit{et al.}~\cite{souza2018open} defined this as silver standard labels. 
        
    \item \textit{Synthetic Images}: Recent studies have proved the generalization ability of a network trained with synthetic images. This means that by creating synthetic datasets using GAN~\cite{thambawita2022singan} or probabilistic models~\cite{billot_synthseg_2021}, your network can predict on real images.
\end{itemize}

  \begin{table}[!h]
\begin{center}
\caption{\label{tab:data_lab} Some examples of automated segmentation tools.}
\begin{tabular}{ c  c }
\textbf{Tool} & \textbf{Structure} \\
\textit{FreeSurfer~\cite{freesurfer}} & {whole brain} \\ 
\textit{FSL~\cite{fsl}} & {Brain tissues} \\ 
\textit{BrainSuite~\cite{brainsuite}} & {skull stripping} \\
\textit{VolBrain~\cite{volbrain}} & {subcortical}\\ 
\textit{SLANT~\cite{huo20193d}} & {whole brain} \\ 
\textit{SynthSeg~\cite{synthseg}} & {whole brain}  \\ 
\textit{FastSurfer~\cite{fastsurfer}} & {whole brain} \\ 
\textit{QuickNAT~\cite{quicknat}} & {whole brain}\\ 
\textit{DeepNAT~\cite{deepnat}} & {whole brain}  \\ 
\textit{Pulmonary Toolkit~\footnote{\url{https://github.com/tomdoel/pulmonarytoolkit}}} & {lung}\\
\textit{Lungmask~\cite{hofmanninger2020automatic}} & {lung}\\
\textit{MEDSeg~\cite{carmo2021multitasking}} & {airway, lung and findings} \\
\end{tabular} 
\end{center}
\end{table}

\subsubsection{Checking the data}

You defined your problem, talked to physicians, and understood what data you need. Now it is time to look at it. Once again, 
It may sound obvious, however, if you are working with medical imaging, you must visually inspect the images before training. 
There are three moments during your application development when you need to look at a sample of the data.
First, before any preprocessing. Check your raw data, see if the labels are coherent and if the images follow the same orientation acquisition. If you are dealing with diverse datasets, check the difference between them. Use this time to understand which preprocessing algorithms you will need to run on your image and to analyze if you understand the structure(s) you are dealing with. You may use visualization tools such as: ITK-SNAP~\cite{py06nimg}; Freeview~\cite{freesurfer}; 3D Slicer~\cite{kikinis20133d}; or DSI Studio~\cite{yeh2020shape} to see your images; or load them on your python script, using libraries such as: Nibabel~\cite{brett_matthew_2023_7633628}; PyDicom~\footnote{\url{https://pydicom.github.io/}}; SimpleITK~\cite{lowekamp2013design}; or MedPY~\footnote{\url{https://loli.github.io/medpy/}}. If you have doubts about the labels, this is the moment to clarify them, or you may have to train your network again.
Second, look at your data after preprocessing. You must ensure that the network will receive exactly what you intended to deliver. 
Finally, check your predictions according to the input data, if the metrics make sense, or if you need some post-processing algorithm to improve your results.

\subsubsection{Ethics on data}

It is crucial to understand that medical imaging is sensitive data, especially if you deal with patient images and/or pediatric images. To use the data, you need permission from the patients or from their legal representatives. It is imperative to anonymize the data by hiding personal information. Also, it is interesting to run a defacing algorithm in some cases to prevent face reconstruction from 3D images~\cite{diaz2021data}.
If you are using a public dataset, it is important to check if the data attend the ethics committee requirements. You will also need permission from the owner to use the data.


\subsubsection{Preprocessing}
\label{sec:preprocess}

Before inserting the image into the model, several pre-processing steps can be done to the image, such as bias field correction, registration, skull stripping (in the case of brain images), normalization, and clipping.
Registration is required to place multiple images in the same space, especially when working with multi-modality data, and it can be inter or intra-subject. For example, T1-weighted and Diffusion MR images usually have different resolutions and sometimes even acquisition directions. The registration process will transport one of the images to the space of the other, making each voxel have an exact correspondence in both images. A few of the most used tools to perform the registration are FSL, FreeSurfer, and Dipy. They are able to perform linear registration, a method that can basically manipulate the scales and orientation of the images, and nonlinear registration, which can elastically deform the images to adjust for punctual deformations.  
Intensity inhomogeneity, or bias field, is a low-frequency noise that represents intensity variability on a tissue and is caused during the acquisition process of the MR image~\cite{despotovic2010brain}.  Hence, as preprocessing, it is often necessary to apply a bias field correction such as the nonuniform normalization (N3) algorithm~\cite{sled1998nonparametric}. Skull-stripping (SS), on the other hand, is a step that separates brain tissue from the skull in MR images of the head. It is often used when the skull worsens the task results and it is mandatory when comparing brain structure volumes since the volumes need to be normalized by the total intracranial volume. 
When working with CT, clipping usually is a common pre-processing step. It reduces the gray level window to a specific intensity range highlighting desired structures (Fig.~\ref{fig:hu}). 
Finally, the network input is commonly normalized to avoid exploding gradients - usually into the [0,1] or [-1,1] range.

\begin{figure}[ht]
    \centering
    \includegraphics[width=\columnwidth]{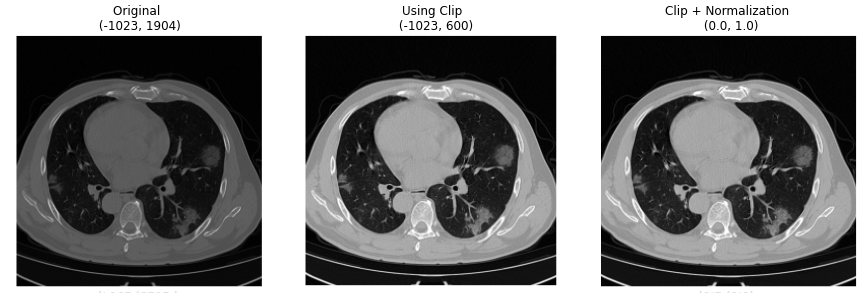}
    \caption{\label{fig:prep}Example of CT image after clipping and normalization. On the left, you can see the original image. On the center, the image after a [-1024,600] clipping: It is possible to see changes in the image contrast. On the right, image after minmax normalization: Visually, there are no changes, however, the minimum and maximum intensity are different.}
\end{figure}

\subsection{Model Training}
 \subsubsection{Data Loading}
    
    After data checking and pre-processing are performed, you should have your data stored in a preprocessed state somewhere, such as preprocessed 3D volumes or preprocessed 2D images. Since medical images are usually multidimensional and large and DL requires a significant amount of data, you will need to pay attention to certain optimization tactics in data loading that are essential for training to be feasible, mainly related to how you save and read those preprocessed files. With some machine learning applications, you could have all the data stored as a variable in RAM, and fit your model there, but that is not viable with the large size of medical images.
    Therefore, your data reading logic will need to save some kind of index to your images and read each item in real-time during training. PyTorch recommends having a Dataset class that manages indexing all your data files and application of data augmentation if necessary and feeding that Dataset class to a Dataloader, which optimizes the loading process using parallelization. The goal in DL training is to use 100\% of your GPU at all times. Therefore, the parallel loading logic implemented by the dataloader is sometimes essential for optimal training speeds. Monitoring your GPU usage during training is a good way to check if your data loading is bottlenecking your training.
    
    There are many ways to optimize storing and reading your pre-processed data, and the exact way this will be done depends on the nature of your data, your hardware environment, and your goals. For example, using compressed formats such as \textit{.npz} can lead to less storage use, but more CPU load in decompressing. Uncompressed \textit{.h5py} files can make organization easier and have less CPU usage, but it can lead to the need for more storage. Even if you have unlimited storage, using uncompressed formats such as \textit{.npy} can be slower due to the slow read speeds of slow hard drives. An example of how to store processed data from CT slices for 2D training using compressed \textit{.npz} is in our case studies (Section~\ref{sec:case}), but should not be taken as the "correct" way. Experimentation must be done in your environment to determine the best storage and reading format for your case. 
    
    \vspace{0.5cm}
    \subsubsection{Data Augmentation}
    
\begin{figure}[ht]
    \centering
    \includegraphics[width=\columnwidth]{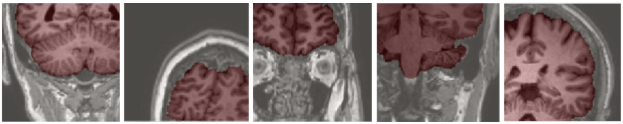}
    \caption{\label{fig:augm}Example of MR images after data augmentation (RandomRotation and RandomCrop). In red, it is possible to see the label overlaying the image.}
    \end{figure}

    When dealing with medical imaging processing, the generalization capacity of the model is of major importance. Once we have several scan manufacturers and different acquisition protocols, each database has its own particularities, especially when compared with data from different medical centers. One could train their model using only one dataset, however, this model would have a loss of performance when applied to data from different centers. 
    Focusing on increasing image variability and improving the generalization of the methods, most authors resort to data augmentation (Fig. \ref{fig:augm}).
    There are several types of augmentation you may use in your method. It is really important to understand which will suit you more or the wrong use of this technique may worsen your results. A few examples of data augmentation focused on medical images are random crop, random rotation, elastic transformation, noise insertion, and intensity transformations (contrast, bright). When dealing with segmentation, it is important to apply the same transforms on the images and labels. Some data augmentation libraries may help, such as Torchvision, Albumentations, and Torch.io.
    
    \vspace{0.5cm}
    \subsubsection{Deep Networks}
    
    Deep Networks are a class of Machine learning methods that are able to extract the features of the data without the need for feature engineering done by a specialist. Instead, the architecture is able to learn not only the task but also the feature extraction on the first convolutional layers.
    
    Usually, for medical images, specialists are focused on the application, not on designing the architecture. Thankfully, there is a large amount of CNN architectures available for general and medical imaging use, in both 2D and 3D formats. Although intuitive, the 3D CNN approach is not always the best performer compared to stacking 2D prediction of each slice. It is important to emphasize that even 2D slices have a thickness in addition to the voxel planar dimension, so the information of each voxel comes from a unity of volume.
    
    We could split the CNN into many different groups, but in this tutorial, we are restricted to two CNN types, one for segmentation and another for classification. Each type of CNN has its characteristics accordingly to the application it is performing. For example, architectures designed for segmentation, usually fully convolutional (every layer is convolutional), are able to generate an output image with the same dimensions as the input. For this type of CNN, the ground truth is composed of masks of the same size as the input images.
    On the other hand, architectures designed for classification present a dense layer after the feature extraction section, and these dense sections have one output for each of the considered classes. In these problems, the ground truth is a single value that represents the class of the input image.
    
    In both types of applications, there are many available architectures. For example, in segmentation, some of the most popular or most performing architectures are the U-Net~\cite{ronneberger2015u}, V-Net~\cite{vnet} and QuickNAT~\cite{roy2019quicknat}. For applications focused on classification, some example architectures are the Resnet~\cite{he2016deep}, Inception~\cite{szegedy2017inception}, DenseNet~\cite{huang2017densely}, MobileNet~\cite{howard2017mobilenets} and EfficientNet~\cite{tan2019efficientnet}.
    \vspace{0.5cm}
    \subsubsection{Loss Function}
    The loss function is defined by a relation between the current output of the model and the desired output, defined by the ground truth. By computing the loss function for the current state of the model, the training framework can define the gradient directions in which the weights are going to be moved to minimize the prediction error of the model.
    
    The loss function must be properly defined depending on the problem the CNN is solving. For segmentation problems, loss functions based on overlap measures are commonly used~\cite{sudre2017generalised}.
    
   It is important to notice that the loss function must be continuous in order to be differentiable, allowing for the optimizer to minimize it, therefore minimizing the error expressed by the loss. However, many problems would require metrics that do not fit this requirement, so the loss function must be properly defined depending on the problem the model is solving. For example, some classification problems search for higher accuracy, which is a non-differentiable function. In this case, a loss based on cross entropy is a valid option. For segmentation problems, a similarity metric such as Dice Coefficient~\cite{sudre2017generalised} is not differentiable in its original set based definition. Therefore, most Dice based losses alter the metric definition to allow for probabilistic inputs to be compared with the binary groundtruth, which also has the benefit of smoothing convergence. This version of Dice is commonly called Dice Loss and can be defined as:

   \begin{equation}
   DiceLoss = 1 - 2\frac{\sum_{i}^{N}p_{i}t_{i}}{\sum_{i}^{N}p_{i}^{2} + \sum_{i}^{N}t_{i}^{2}}
   \label{eq:diceloss}
   \end{equation}
   where $p$ is the probabilistic output of the network between 0 and 1, usually from sigmoid or softmax activations, and $t$ are binary targets. Note that minimizing equation~(\ref{eq:diceloss}) by definition maximizes the value of the Dice metric (Section~\ref{sec:metrics}).
    \vspace{0.5cm}
    \subsubsection{Optimizer}
    
    Neural network training performs adjustments to weights $w$ present in the network, which in the case of a CNN are convolutional kernel values. These weights are changed based on values returned by the loss functions, which measure how "wrong" an output is in relation to a target, which is assumed to be the expected output. The gradient $\delta_{w}$ of a weight represents the change in loss caused by a change in the weights, in other words, a derivation, and guides the training process in the direction of minimizing the loss. The gradient for each weight is commonly calculated using backpropagation in the form of PyTorch's Autograd~\cite{paszke2017automatic}. The optimization process is controlled by optimizers, with the most common ones being Stochastic gradient descent (SGD) and Adaptative momentum estimation (ADAM). 
    
    With the SGD optimizer, the derivation of an updated weight $w_{t}$ for a current discrete time $t$ as a function of past weight $w_{t-1}$ can be expressed as:

    \begin{equation}
    w_{t} = w_{t-1} - \alpha\Delta_{w}{L(O, T)}     
   \end{equation}
    for a loss function $L$ over outputs $O$ and targets $T$. 
    
    SGD can also be implemented with momentum, where past weights are also taken into consideration. The learning rate (LR) controls the speed of the optimization process. Finding the correct learning rate is a key factor in training a CNN. A high learning rate can make the model skip states where the minimum loss is achieved, whereas a low learning rate can lead to very slow convergence, and to the model being stuck in a local minimum. Some learning rate schedulers change the learning rate during training with a function of the number of epochs passed. Some other optimizers try to change the learning rate adaptively, such as ADAM. ADAM computes a learning rate per parameter, instead of using a global learning rate, and takes into consideration a moving average of gradients. More details can be found in its original publication in~\cite{kingma2014adam}. ADAM and its variations~\cite{loshchilov2017decoupled} are a good initial choice of optimizer due to their adaptability.
    A weight update for ADAM can be expressed as:

    \begin{equation}
    w_{t} = w_{t-1} - \eta\frac{m_{t}}{\sqrt{v_{t}} + \epsilon}     
    \end{equation}
   where $\eta$ is the step-size, that can vary between iterations. $m_{t}$ and $v_{t}$ are bias corrected first and second momentums, respectively. These momentums are a function of gradients and the square of the gradients, also in relation to an input and loss function. More details can be found in its original publication in~\cite{kingma2014adam}. The usage of momentum and moving averages avoids the next step to be completely determined by the current batch of data since batches can be very randomized. The momentum keeps pointing to a general direction of minimization, where the current gradient points to the minimization for the current batch. The final step can be defined by a combination of the two.
   
    
    
    
    
    
\subsection{Post processing}

\begin{figure}[ht]
    \centering
    \includegraphics[width=\columnwidth]{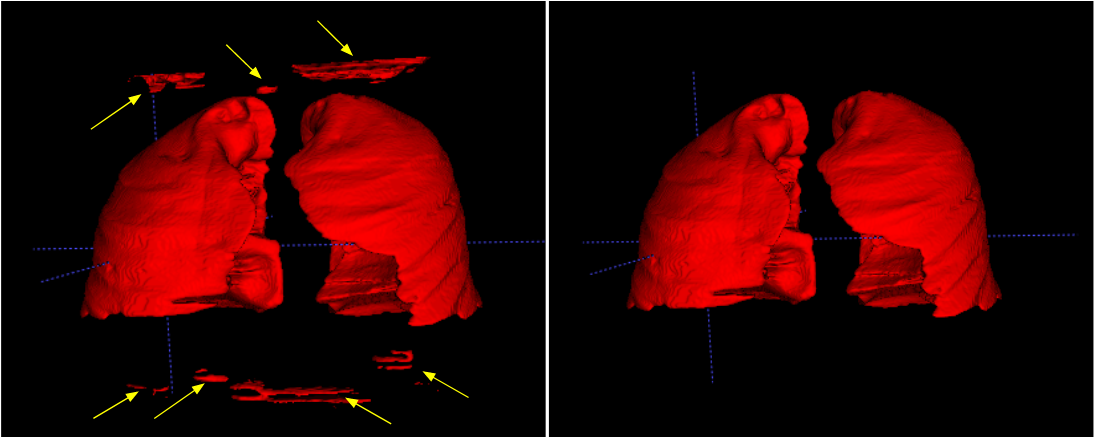}
    \caption{\label{fig:post}Prediction noise pointed by yellow arrows (left image) can be reduced (right image) using post-processing algorithms such as CC filters.}
\end{figure}

Post-processing is the process of editing the network output to enhance the results. There are several post-processing options based on the end result you want to achieve. For instance, the analysis of connected components (CC) over the three dimensions. By checking the CC on a segmentation output, it is possible to remove miss-classified voxels (Fig.~\ref{fig:post}).
The threshold analysis is also used to improve a network outcome. It needs to be done on the validation set and the main idea is to find a threshold that better improves the final result. For this matter, it is necessary to vary the thresholding by analyzing the metric you want to improve. This may be applied to different applications such as classification and segmentation.
Seung-lab~\footnote{\url{https://github.com/seung-lab/connected-components-3d}} provides a useful 3D connected components generation method which supports images containing many different labels, not just binary images. It also supports continuously valued images such as grayscale microscope images with an algorithm that joins together nearby values. The benefit of this package is that it labels all connected components in one shot, improving performance by one or more orders of magnitude.

\subsection{Evaluation Metrics}
\label{sec:metrics}
Concerning segmentation models, different metrics should be used to evaluate results due to their complementarity in expressing how well the method's output is similar to the ground truth~\cite{taha2015metrics}. Some examples include the Dice coefficient (Dice), Hausdorff Distance (HD), Hausdorff Average Distance (AVD), and Volume Simmetry, among others.
Considering \textbf{A} as the model prediction and \textbf{M} as the label, we may define:

\vspace{0.5cm}
\subsubsection{Dice Coefficient}

The \textit{DC} is an overlap measure defined as follows:
\begin{equation}
    DC = \frac{2*\left | M\cap A \right |}{|M|+ |A|}
\end{equation}

\textit{DC} is sensitive to small segmentation and does not identify boundary errors. However, it can be used as a measure of reproducibility and is widely used for medical imaging segmentation analysis, being the most used metric in the medical imaging segmentation field \cite{taha2015metrics}. 
\textit{DC} results may be in the [0,1] range, where 1 a perfect \textit{DC}
\vspace{0.5cm}
\subsubsection{Hausdorff Distance}

The \textit{HD} measures the distance between two sets of points.  

\begin{equation}
    HD(A,M) = max(h(A,M),h(M,A))
\end{equation}
where:

\begin{equation}
        h(A,M) = \max_{a\epsilon A}\min_{m\epsilon M}\left \| a-m \right \|
\end{equation}
\vspace{0.5cm}
\subsubsection{Average Hausdorff Distance}
    
As the name suggests, \textit{AVD} is the averaged Hausdorff Distance over all points.

\begin{equation}
    AVD(A, M) = max(d(A, M), d(M,A))
\end{equation}
where:

\begin{equation}
  d(A,M) = \frac{1}{N}\sum_{a\epsilon A}\min_{m\epsilon M}\left \| a-m \right \|
\end{equation}

Similar to \textit{HD}, it is also a spatial distance metric that is robust to small structures. However, on average, it is less sensitive to outliers~\cite{taha2015metrics}. The smaller the \textit{AVD} between manual and automated segmentation, the better the automated segmentation.

\vspace{0.5cm}
\subsubsection{Volume Similarity}

Finally, the VS calculates the similarity between the two samples. 

\begin{equation}
   VS = 1 -  \frac{\left | \left | A  \right |-\left |  M \right | \right |}{\left | A \right |+\left |  M \right |}
\end{equation}
being $ \left | X  \right |$ the module of X. Although it ignores borders and overlaps, \textit{VS} is a good metric for analyzing the segmentation volume when determining the volume of the structure is the main goal~\cite{taha2015metrics}. 
\textit{VS} results may be in a [0,1] range, being 1 a perfect \textit{VS}

\section{Statistic Analysis}
Statistical analysis is an important task to analyze the reliability of the obtained results obtained. The main idea is to compare groups and verify if they are statistically equivalent or different, thus obtaining a significant answer or not for the problem in question.

At first, it is important to stipulate which hypotheses are being tested. In a hypothesis test, the Null Hypothesis ($H_{0}$) is the equality hypothesis, which argues that the groups being analyzed are statistically equal. The alternative hypothesis, ($H_{1}$), is the complement of the null hypothesis and states that the effect between the groups under study exists, that is, that the groups are statistically different.

A statistical test can be one-tailed or two-tailed. In the two-tailed test, the alternative hypothesis is a hypothesis of inequality ($\neq$), not taking into account if values are less or greater ($<$ or $>$).

There are several statistical tests and the correct choice will depend on the analysis you want to perform. Our focus will be to describe step by step how to perform a statistical test for two groups with quantitative response variables, which are the most used analyses in medical data.

\subsection{Check if the groups are independent or dependent}

Two groups are independent (unpaired) if the sample selected from one of the populations is not related to the sample selected from the second population. For example, if we want to compare patients $\times$ control.

Two groups are dependent (paired) if each member in a sample correspondent with a member from another sample. For example, collected data from the same cohort, before and after treatment.

\subsection{Check data normality}

To perform a confident analysis, it is important to use the statistical test that best represents the data, so before choosing the test it is important to check the distribution of the data.
Normality is a characteristic of the data in which the majority (higher frequency) of the sample values are close to the mean value of all samples. Normality can be visualized through the scatter histogram and the box-plot plot. The data are normally distributed if their scatter histogram is bell-shaped. A boxplot with many outliers is a characteristic of data without distribution, that is, of non-parametric data. However, it is often difficult to define normality just by visualizing the data. 
 According to~\cite{razali2011power}, the most powerful statistical test to verify data normality is the Shapiro Wilk test. The Shapiro Wilk test uses the following hypothesis:

$H_{0}$ : data distribution = normal $\rightarrow$ $p > \alpha$

$H_{1}$ : data distribution $\neq$ normal  $\rightarrow$ $p \leq \alpha$
where alpha is the significance level stipulated as a priory, a limit established for the rejection or note null hypothesis based on the value of \textit{p}. In the medical area, it is often used  $\alpha=0.05$.

\subsection{Use the appropriate statistical test to compare two groups}

Parametric tests require a normal data distribution, while non-parametric tests do not require normal distribution.


\subsubsection{Independent \textit{t} test:}

The t-test for two independent samples (also called independent Student's t-test), allows comparing means between two independent groups. The dependent variable is numerical with normal distribution and the independent variable is categorical with two categories. Before performing the independent t-test, it is important to verify whether the variances of the groups are homogeneous or not using the Levene test~\cite{gastwirth2009impact}.
The hypotheses of the independent t-test are:

$H_{0}$ : mean of the groups are same $\rightarrow$ $p > \alpha$

$H_{1}$ : mean of the groups are different $\rightarrow$ $p \leq \alpha$

\subsubsection{Mann-Whitney test:}

Mann-Whitney test compares two independent groups. It is a non-parametric alternative to the independent t-test. The Mann-Whitney test is used for data that do not have a normal distribution. That is, the mean is not a good representation of the data set and the median is the measure that best represents the data \cite{nachar2008mann}.
The test hypotheses are:

$H_{0}$ : median of the groups are same $\rightarrow$ $p > \alpha$

$H_{1}$ : median of the groups are different $\rightarrow$ $p \leq \alpha$

An important observation is that we can find the Mann-Whitey test as a test that compares the medians, but deep down, it is not just comparing the medians of the two groups, it is comparing the distributions \cite{hart2001mann}. It may happen that the same median is found in both groups under study and the Mann-Whitney test is significant, that is, this significance indicates that there is a difference in the distributions.

\subsubsection{Paired \textit{t} test:}

Paired Student's t-test is a type of hypothesis test that allows comparing the mean of two paired groups. The dependent variable must be numerical, and the independent variable is composed of two paired groups. To use this test the data must be represented by a normal or approximately normal distribution. The paired t-test hypotheses are the same as the independent t-test.

\subsubsection{Wilcoxon test:}

The Wilcoxon test, also known as the Wilcoxon signed-rank test, is based on ranks and allows comparing two paired samples\cite{king2018statistical}. It is a nonparametric test corresponding to the paired t-test. This method considers the size of the differences in the case under study. The Wilcoxon signed-rank test is used to test differences in population. They are generally used for data that do not have a normal distribution. This type of data is rarely well represented by the mean and the median is the measure of central tendency that best represents the data. The hypotheses of this test are the same as in the Mann Whitney test.

\section{Case Studies}
\label{sec:case}
We have prepared a practical demonstration of most principles showcased in this paper for a CT and an MR image case using public data ~\cite{antonelli2022medical, ma_jun_2020_3757476} in the following repository: \url{https://github.com/MICLab-Unicamp/Medical-Imaging-Tutorial}. 

\section{Conclusion}

In this manuscript we have gone through a brief introduction to the field of medical imaging analysis with deep learning, focusing on segmentation of images. A workflow guidance is provided, with tips and good practices for all phases commonly present in this field of research. In addition, we provide hands-on examples using public data, which will be used in our tutorial session.

\section*{Acknowledgment}

We thank MICLab students Álvaro Capelo, Beatriz Vicente, Bruno Santos, Gabriel Dias, Jean Ribeiro, and Joany Rodrigues, for participating in our internal seminars. Livia Rodrigues and Thays Abreu thanks the Higher Education Personnel Improvement Coordination (CAPES). Diedre Carmo thanks grant \#2019/21964-4, São Paulo Research Foundation (FAPESP). Gustavo Pinheiro, Roberto Lotufo, and Leticia Rittner thank the National Scientific and Technological Development Council (CNPq).




\bibliographystyle{IEEEtran} 
\bibliography{bib}

\begin{thebibliography}{100}
\providecommand{\url}[1]{#1}
\csname url@samestyle\endcsname
\providecommand{\newblock}{\relax}
\providecommand{\bibinfo}[2]{#2}
\providecommand{\BIBentrySTDinterwordspacing}{\spaceskip=0pt\relax}
\providecommand{\BIBentryALTinterwordstretchfactor}{4}
\providecommand{\BIBentryALTinterwordspacing}{\spaceskip=\fontdimen2\font plus
\BIBentryALTinterwordstretchfactor\fontdimen3\font minus
  \fontdimen4\font\relax}
\providecommand{\BIBforeignlanguage}[2]{{%
\expandafter\ifx\csname l@#1\endcsname\relax
\typeout{** WARNING: IEEEtran.bst: No hyphenation pattern has been}%
\typeout{** loaded for the language `#1'. Using the pattern for}%
\typeout{** the default language instead.}%
\else
\language=\csname l@#1\endcsname
\fi
#2}}
\providecommand{\BIBdecl}{\relax}
\BIBdecl

\bibitem{lauterbur1973image}
P.~C. Lauterbur, ``Image formation by induced local interactions: examples
  employing nuclear magnetic resonance,'' \emph{nature}, vol. 242, no. 5394,
  pp. 190--191, 1973.

\bibitem{edelman2014history}
R.~R. Edelman, ``The history of {MR} imaging as seen through the pages of
  radiology,'' \emph{Radiology}, vol. 273, no.~2S, pp. S181--S200, 2014.

\bibitem{kaul2020history}
V.~Kaul, S.~Enslin, and S.~A. Gross, ``History of artificial intelligence in
  medicine,'' \emph{Gastrointestinal endoscopy}, vol.~92, no.~4, pp. 807--812,
  2020.

\bibitem{aurna2022classification}
N.~F. Aurna, M.~A. Yousuf, K.~A. Taher, A.~Azad, and M.~A. Moni, ``A
  classification of {MRI} brain tumor based on two stage feature level ensemble
  of deep cnn models,'' \emph{Computers in Biology and Medicine}, vol. 146, p.
  105539, 2022.

\bibitem{deepak2019brain}
S.~Deepak and P.~Ameer, ``Brain tumor classification using deep {CNN} features
  via transfer learning,'' \emph{Computers in biology and medicine}, vol. 111,
  p. 103345, 2019.

\bibitem{carmo2021rapidly}
D.~Carmo, I.~Campiotti, L.~Rodrigues, I.~Fantini, G.~Pinheiro, D.~Moraes,
  R.~Nogueira, L.~Rittner, and R.~Lotufo, ``Rapidly deploying a covid-19
  decision support system in one of the largest brazilian hospitals,''
  \emph{Health Informatics Journal}, vol.~27, no.~3, p. 14604582211033017,
  2021.

\bibitem{gauriau2015multi}
R.~Gauriau, R.~Cuingnet, D.~Lesage, and I.~Bloch, ``Multi-organ localization
  with cascaded global-to-local regression and shape prior,'' \emph{Medical
  image analysis}, vol.~23, no.~1, pp. 70--83, 2015.

\bibitem{noothout2020deep}
J.~M. Noothout, B.~D. De~Vos, J.~M. Wolterink, E.~M. Postma, P.~A. Smeets,
  R.~A. Takx, T.~Leiner, M.~A. Viergever, and I.~I{\v{s}}gum, ``Deep
  learning-based regression and classification for automatic landmark
  localization in medical images,'' \emph{IEEE transactions on medical
  imaging}, vol.~39, no.~12, pp. 4011--4022, 2020.

\bibitem{beauferris2022multi}
Y.~Beauferris, J.~Teuwen, D.~Karkalousos, N.~Moriakov, M.~Caan, G.~Yiasemis,
  L.~Rodrigues, A.~Lopes, H.~Pedrini, L.~Rittner \emph{et~al.}, ``Multi-coil
  mri reconstruction challenge—assessing brain mri reconstruction models and
  their generalizability to varying coil configurations,'' \emph{Frontiers in
  Neuroscience}, vol.~16, p. 919186, 2022.

\bibitem{pinheiro2018v}
G.~R. Pinheiro, R.~Voltoline, M.~Bento, and L.~Rittner, ``V-net and u-net for
  ischemic stroke lesion segmentation in a small dataset of perfusion data,''
  in \emph{International MICCAI Brainlesion Workshop}.\hskip 1em plus 0.5em
  minus 0.4em\relax Springer, 2018, pp. 301--309.

\bibitem{carmo2021multitasking}
D.~Carmo, I.~Campiotti, I.~Fantini, L.~Rodrigues, L.~Rittner, and R.~Lotufo,
  ``Multitasking segmentation of lung and covid-19 findings in ct scans using
  modified efficientdet, unet and mobilenetv3 models,'' in \emph{17th
  International Symposium on Medical Information Processing and Analysis}, vol.
  12088.\hskip 1em plus 0.5em minus 0.4em\relax SPIE, 2021, pp. 65--74.

\bibitem{roy2019quicknat}
A.~G. Roy, S.~Conjeti, N.~Navab, C.~Wachinger, A.~D.~N. Initiative
  \emph{et~al.}, ``{QuickNAT}: A fully convolutional network for quick and
  accurate segmentation of neuroanatomy,'' \emph{NeuroImage}, vol. 186, pp.
  713--727, 2019.

\bibitem{ronneberger2015u}
O.~Ronneberger, P.~Fischer, and T.~Brox, ``U-net: Convolutional networks for
  biomedical image segmentation,'' in \emph{International Conference on Medical
  image computing and computer-assisted intervention}.\hskip 1em plus 0.5em
  minus 0.4em\relax Springer, 2015, pp. 234--241.

\bibitem{carmo2021hippocampus}
D.~Carmo, B.~Silva, C.~Yasuda, L.~Rittner, R.~Lotufo, A.~D.~N. Initiative
  \emph{et~al.}, ``Hippocampus segmentation on epilepsy and alzheimer's disease
  studies with multiple convolutional neural networks,'' \emph{Heliyon},
  vol.~7, no.~2, p. e06226, 2021.

\bibitem{rodrigues2020hypothalamus}
L.~Rodrigues, T.~Rezende, A.~Zanesco, A.~L. Hernandez, M.~Franca, and
  L.~Rittner, ``Hypothalamus fully automatic segmentation from mr images using
  a u-net based architecture,'' in \emph{15th International Symposium on
  Medical Information Processing and Analysis}, vol. 11330.\hskip 1em plus
  0.5em minus 0.4em\relax SPIE, 2020, pp. 144--150.

\bibitem{rahman2020automatic}
M.~M. Rahman, L.~Duerselen, and A.~M. Seitz, ``Automatic segmentation of knee
  menisci - {A} systematic review,'' \emph{Artificial Intelligence in
  Medicine}, vol. 105, p. 101849, 2020.

\bibitem{tang2020two}
W.~Tang, D.~Zou, S.~Yang, J.~Shi, J.~Dan, and G.~Song, ``A two-stage approach
  for automatic liver segmentation with faster {R-CNN and DeepLab},''
  \emph{Neural Computing and Applications}, vol.~32, no.~11, pp. 6769--6778,
  2020.

\bibitem{buzug2011computed}
T.~M. Buzug, ``Computed tomography,'' in \emph{Springer handbook of medical
  technology}.\hskip 1em plus 0.5em minus 0.4em\relax Springer, 2011, pp.
  311--342.

\bibitem{suzuki2017overview}
K.~Suzuki, ``Overview of deep learning in medical imaging,'' \emph{Radiological
  physics and technology}, vol.~10, no.~3, pp. 257--273, 2017.

\bibitem{kilsdonk2014improved}
I.~D. Kilsdonk, M.~P. Wattjes, A.~Lopez-Soriano, J.~Kuijer, M.~C. de~Jong,
  W.~L. de~Graaf, M.~Conijn, C.~H. Polman, P.~R. Luijten, J.~J. Geurts
  \emph{et~al.}, ``Improved differentiation between {MS} and vascular brain
  lesions using {FLAIR}* at 7 tesla,'' \emph{European radiology}, vol.~24,
  no.~4, pp. 841--849, 2014.

\bibitem{basser1995diffusion}
P.~Basser and C.~Pierpaoli, ``Diffusion tensor {MRI}: A new tool for
  elucidating tissue microstructure and organization,''
  \emph{ASME-PUBLICATIONS-BED}, vol.~29, pp. 367--367, 1995.

\bibitem{huettel2004functional}
S.~A. Huettel, A.~W. Song, G.~McCarthy \emph{et~al.}, \emph{Functional magnetic
  resonance imaging}.\hskip 1em plus 0.5em minus 0.4em\relax Sinauer Associates
  Sunderland, 2004, vol.~1.

\bibitem{harris1986nuclear}
R.~K. Harris, \emph{Nuclear magnetic resonance spectroscopy}.\hskip 1em plus
  0.5em minus 0.4em\relax John Wiley and Sons Inc., New York, NY, 1986.

\bibitem{carmo2022systematic}
D.~Carmo, J.~Ribeiro, S.~Dertkigil, S.~Appenzeller, R.~Lotufo, and L.~Rittner,
  ``A systematic review of automated segmentation methods and public datasets
  for the lung and its lobes and findings on computed tomography images,''
  \emph{Yearbook of Medical Informatics}, vol.~31, no.~01, pp. 277--295, 2022.

\bibitem{lecun2015deep}
Y.~LeCun, Y.~Bengio, and G.~Hinton, ``Deep learning,'' \emph{nature}, vol. 521,
  no. 7553, pp. 436--444, 2015.

\bibitem{sudre2017generalised}
C.~H. Sudre, W.~Li, T.~Vercauteren, S.~Ourselin, and M.~Jorge~Cardoso,
  ``Generalised dice overlap as a deep learning loss function for highly
  unbalanced segmentations,'' in \emph{Deep learning in medical image analysis
  and multimodal learning for clinical decision support}.\hskip 1em plus 0.5em
  minus 0.4em\relax Springer, 2017, pp. 240--248.

\bibitem{DBLP:journals/corr/KingmaB14}
\BIBentryALTinterwordspacing
D.~P. Kingma and J.~Ba, ``Adam: A method for stochastic optimization,'' in
  \emph{ICLR (Poster)}, 2015. [Online]. Available:
  \url{http://arxiv.org/abs/1412.6980}
\BIBentrySTDinterwordspacing

\bibitem{DBLP:journals/corr/IoffeS15}
\BIBentryALTinterwordspacing
S.~Ioffe and C.~Szegedy, ``Batch normalization: Accelerating deep network
  training by reducing internal covariate shift,'' \emph{CoRR}, vol.
  abs/1502.03167, 2015. [Online]. Available:
  \url{http://arxiv.org/abs/1502.03167}
\BIBentrySTDinterwordspacing

\bibitem{xu2015empirical}
B.~Xu, N.~Wang, T.~Chen, and M.~Li, ``Empirical evaluation of rectified
  activations in convolutional network,'' \emph{arXiv preprint
  arXiv:1505.00853}, 2015.

\bibitem{shamshad2022transformers}
F.~Shamshad, S.~Khan, S.~W. Zamir, M.~H. Khan, M.~Hayat, F.~S. Khan, and H.~Fu,
  ``Transformers in medical imaging: A survey,'' \emph{arXiv preprint
  arXiv:2201.09873}, 2022.

\bibitem{he2016deep}
K.~He, X.~Zhang, S.~Ren, and J.~Sun, ``Deep residual learning for image
  recognition,'' in \emph{Proceedings of the IEEE conference on computer vision
  and pattern recognition}, 2016, pp. 770--778.

\bibitem{tan2019efficientnet}
M.~Tan and Q.~Le, ``Efficientnet: Rethinking model scaling for convolutional
  neural networks,'' in \emph{International conference on machine
  learning}.\hskip 1em plus 0.5em minus 0.4em\relax PMLR, 2019, pp. 6105--6114.

\bibitem{liu2022convnet}
Z.~Liu, H.~Mao, C.-Y. Wu, C.~Feichtenhofer, T.~Darrell, and S.~Xie, ``A convnet
  for the 2020s,'' in \emph{Proceedings of the IEEE/CVF Conference on Computer
  Vision and Pattern Recognition}, 2022, pp. 11\,976--11\,986.

\bibitem{chen2018encoder}
L.-C. Chen, Y.~Zhu, G.~Papandreou, F.~Schroff, and H.~Adam, ``Encoder-decoder
  with atrous separable convolution for semantic image segmentation,'' in
  \emph{Proceedings of the European conference on computer vision (ECCV)},
  2018, pp. 801--818.

\bibitem{abdollahi2020vnet}
A.~Abdollahi, B.~Pradhan, and A.~Alamri, ``Vnet: An end-to-end fully
  convolutional neural network for road extraction from high-resolution remote
  sensing data,'' \emph{IEEE Access}, vol.~8, pp. 179\,424--179\,436, 2020.

\bibitem{karwoski2008processing}
R.~A. Karwoski, B.~Bartholmai, V.~A. Zavaletta, D.~Holmes, and R.~A. Robb,
  ``Processing of ct images for analysis of diffuse lung disease in the lung
  tissue research consortium,'' in \emph{Medical imaging 2008: physiology,
  function, and structure from medical images}, vol. 6916.\hskip 1em plus 0.5em
  minus 0.4em\relax SPIE, 2008, pp. 356--364.

\bibitem{armato2011rd}
S.~Armato, ``rd, mclennan g, bidaut l, mcnitt-gray mf, meyer cr, reeves ap, et
  al. the lung image database consortium (lidc) and image database resource
  initiative (idri): A completed reference database of lung nodules on ct
  scans,'' \emph{Med Phys}, vol.~38, no.~2, pp. 915--31, 2011.

\bibitem{lola11}
``{LOLA11},'' \url{https://lola11.grand-challenge.org/}.

\bibitem{luna16}
``{LUNA16},'' \url{https://luna16.grand-challenge.org/.}

\bibitem{yan2020ccap}
T.~Yan, ``Ccap,'' \emph{IEEE Dataport}, 2020.

\bibitem{msl}
``{MSL},''
  \url{http://medicalsegmentation.com/covid19/,https://www.medseg.ai/.}

\bibitem{morozov2020mosmeddata}
S.~P. Morozov, A.~Andreychenko, N.~Pavlov, A.~Vladzymyrskyy, N.~Ledikhova,
  V.~Gombolevskiy, I.~A. Blokhin, P.~Gelezhe, A.~Gonchar, and V.~Y. Chernina,
  ``Mosmeddata: Chest ct scans with covid-19 related findings dataset,''
  \emph{arXiv preprint arXiv:2005.06465}, 2020.

\bibitem{jun2020covid}
M.~Jun, G.~Cheng, W.~Yixin, A.~Xingle, G.~Jiantao, Y.~Ziqi, Z.~Minqing, L.~Xin,
  D.~Xueyuan, C.~Shucheng \emph{et~al.}, ``Covid-19 ct lung and infection
  segmentation dataset. 2020,'' 2020.

\bibitem{kumar2018segmentation}
A.~Kumar, S.~Agarwala, A.~K. Dhara, D.~Nandi, S.~B. Thakur, A.~K. Bhadra, and
  A.~Sadhu, ``Segmentation of lung field in hrct images using u-net based fully
  convolutional networks,'' in \emph{Annual Conference on Medical Image
  Understanding and Analysis}.\hskip 1em plus 0.5em minus 0.4em\relax Springer,
  2018, pp. 84--93.

\bibitem{murphy2011evaluation}
K.~Murphy, B.~Van~Ginneken, J.~M. Reinhardt, S.~Kabus, K.~Ding, X.~Deng,
  K.~Cao, K.~Du, G.~E. Christensen, V.~Garcia \emph{et~al.}, ``Evaluation of
  registration methods on thoracic ct: the empire10 challenge,'' \emph{IEEE
  transactions on medical imaging}, vol.~30, no.~11, pp. 1901--1920, 2011.

\bibitem{vessel12}
``{VESSEL12},'' \url{https://vessel12.grand-challenge.org.}

\bibitem{langs2012visceral}
G.~Langs, A.~Hanbury, B.~Menze, and H.~M{\"u}ller, ``Visceral: towards large
  data in medical imaging—challenges and directions,'' in \emph{MICCAI
  international workshop on medical content-based retrieval for clinical
  decision support}.\hskip 1em plus 0.5em minus 0.4em\relax Springer, 2012, pp.
  92--98.

\bibitem{dsb}
``{DSB},'' \url{https://www.kaggle.com/c/data-science-bowl-2017.}

\bibitem{kaggle}
``{Kaggle},'' \url{https://www.kaggle.com/kmader/finding-lungs-in-ct-data.}

\bibitem{wang2016automatic}
J.~Wang and H.~Guo, ``Automatic approach for lung segmentation with
  juxta-pleural nodules from thoracic ct based on contour tracing and
  correction,'' \emph{Computational and mathematical methods in medicine}, vol.
  2016, 2016.

\bibitem{lo2012extraction}
P.~Lo, B.~Van~Ginneken, J.~M. Reinhardt, T.~Yavarna, P.~A. De~Jong, B.~Irving,
  C.~Fetita, M.~Ortner, R.~Pinho, J.~Sijbers \emph{et~al.}, ``Extraction of
  airways from ct (exact'09),'' \emph{IEEE Transactions on Medical Imaging},
  vol.~31, no.~11, pp. 2093--2107, 2012.

\bibitem{cid2015efficient}
Y.~D. Cid, O.~A.~J. Del~Toro, A.~Depeursinge, and H.~M{\"u}ller, ``Efficient
  and fully automatic segmentation of the lungs in ct volumes.'' in
  \emph{VISCERAL Challenge@ ISBI}, 2015, pp. 31--35.

\bibitem{soares2020explainable}
E.~Soares, P.~Angelov, S.~Biaso, M.~H. Froes, and D.~K. Abe,
  ``Explainable-by-design approach for covid-19 classification via ct-scan,''
  2020.

\bibitem{tcia}
``{TCIA},''
  \url{https://wiki.cancerimagingarchive.net/display/Public/CT+Images+in+COVID-19.}

\bibitem{jack2008alzheimer}
C.~R. Jack~Jr, M.~A. Bernstein, N.~C. Fox, P.~Thompson, G.~Alexander,
  D.~Harvey, B.~Borowski, P.~J. Britson, J.~L.~Whitwell, C.~Ward \emph{et~al.},
  ``The alzheimer's disease neuroimaging initiative (adni): Mri methods,''
  \emph{Journal of Magnetic Resonance Imaging: An Official Journal of the
  International Society for Magnetic Resonance in Medicine}, vol.~27, no.~4,
  pp. 685--691, 2008.

\bibitem{lamontagne2019oasis}
P.~J. LaMontagne, T.~L. Benzinger, J.~C. Morris, S.~Keefe, R.~Hornbeck,
  C.~Xiong, E.~Grant, J.~Hassenstab, K.~Moulder, A.~Vlassenko \emph{et~al.},
  ``{OASIS-3: longitudinal neuroimaging, clinical, and cognitive dataset for
  normal aging and Alzheimer disease},'' \emph{{MedRxiv}}, 2019.

\bibitem{van2012human}
D.~C. Van~Essen, K.~Ugurbil, E.~Auerbach, D.~Barch, T.~E. Behrens, R.~Bucholz,
  A.~Chang, L.~Chen, M.~Corbetta, S.~W. Curtiss \emph{et~al.}, ``The human
  connectome project: a data acquisition perspective,'' \emph{Neuroimage},
  vol.~62, no.~4, pp. 2222--2231, 2012.

\bibitem{IXI}
``{IXI Dataset},'' \url{https://brain-development.org/ixi-dataset/}, accessed:
  2021-06-12.

\bibitem{di2014autism}
A.~Di~Martino, C.-G. Yan, Q.~Li, E.~Denio, F.~X. Castellanos, K.~Alaerts, J.~S.
  Anderson, M.~Assaf, S.~Y. Bookheimer, M.~Dapretto \emph{et~al.}, ``The autism
  brain imaging data exchange: towards a large-scale evaluation of the
  intrinsic brain architecture in autism,'' \emph{Molecular psychiatry},
  vol.~19, no.~6, pp. 659--667, 2014.

\bibitem{landman2011multi}
B.~A. Landman, A.~J. Huang, A.~Gifford, D.~S. Vikram, I.~A.~L. Lim, J.~A.
  Farrell, J.~A. Bogovic, J.~Hua, M.~Chen, S.~Jarso \emph{et~al.},
  ``Multi-parametric neuroimaging reproducibility: a 3-{T} resource study,''
  \emph{Neuro{I}mage}, vol.~54, no.~4, pp. 2854--2866, 2011.

\bibitem{frisoni2015eadc}
G.~B. Frisoni, C.~R. Jack~Jr, M.~Bocchetta, C.~Bauer, K.~S. Frederiksen,
  Y.~Liu, G.~Preboske, T.~Swihart, M.~Blair, E.~Cavedo \emph{et~al.}, ``The
  eadc-adni harmonized protocol for manual hippocampal segmentation on magnetic
  resonance: Evidence of validity,'' \emph{Alzheimer's \& Dementia}, vol.~11,
  no.~2, pp. 111--125, 2015.

\bibitem{souza2018open}
R.~Souza, O.~Lucena, J.~Garrafa, D.~Gobbi, M.~Saluzzi, S.~Appenzeller,
  L.~Rittner, R.~Frayne, and R.~Lotufo, ``An open, multi-vendor,
  multi-field-strength brain mr dataset and analysis of publicly available
  skull stripping methods agreement,'' \emph{NeuroImage}, vol. 170, pp.
  482--494, 2018.

\bibitem{brats}
B.~H. Menze, A.~Jakab, S.~Bauer, J.~Kalpathy-Cramer, K.~Farahani, J.~Kirby,
  Y.~Burren, N.~Porz, J.~Slotboom, R.~Wiest \emph{et~al.}, ``The multimodal
  brain tumor image segmentation benchmark (brats),'' \emph{IEEE transactions
  on medical imaging}, vol.~34, no.~10, pp. 1993--2024, 2014.

\bibitem{rodrigues2022benchmark}
L.~Rodrigues, T.~Rezende, G.~Wertheimer, Y.~Santos, M.~Franca, and L.~Rittner,
  ``A benchmark for hypothalamus segmentation on t1-weighted mr images,''
  \emph{NeuroImage}, p. 119741, 2022.

\bibitem{shattuck2008construction}
D.~W. Shattuck, M.~Mirza, V.~Adisetiyo, C.~Hojatkashani, G.~Salamon, K.~L.
  Narr, R.~A. Poldrack, R.~M. Bilder, and A.~W. Toga, ``Construction of a 3d
  probabilistic atlas of human cortical structures,'' \emph{Neuroimage},
  vol.~39, no.~3, pp. 1064--1080, 2008.

\bibitem{puccio2016preprocessed}
B.~Puccio, J.~P. Pooley, J.~S. Pellman, E.~C. Taverna, and R.~C. Craddock,
  ``The preprocessed connectomes project repository of manually corrected
  skull-stripped t1-weighted anatomical mri data,'' \emph{Gigascience}, vol.~5,
  no.~1, pp. s13\,742--016, 2016.

\bibitem{ibsr}
J.~Frazier, V.~Caviness, D.~Kennedy, A.~Worth, C.~Haselgrove, D.~Caplan, and
  N.~Makris, ``Internet brain segmentation repository (ibsr) 1.5 mm dataset,''
  \emph{Collections}, vol.~10, no. C60W3M, p. C6RC85, 1910.

\bibitem{buda2019association}
M.~Buda, A.~Saha, and M.~A. Mazurowski, ``Association of genomic subtypes of
  lower-grade gliomas with shape features automatically extracted by a deep
  learning algorithm,'' \emph{Computers in biology and medicine}, vol. 109, pp.
  218--225, 2019.

\bibitem{sivaswamy2021sub}
J.~Sivaswamy, A.~J. Thottupattu, R.~Mehta, R.~Sheelakumari, C.~Kesavadas
  \emph{et~al.}, ``Sub-cortical structure segmentation database for young
  population,'' \emph{arXiv preprint arXiv:2111.01561}, 2021.

\bibitem{thambawita2022singan}
V.~Thambawita, P.~Salehi, S.~A. Sheshkal, S.~A. Hicks, H.~L. Hammer, S.~Parasa,
  T.~d. Lange, P.~Halvorsen, and M.~A. Riegler, ``Singan-seg: Synthetic
  training data generation for medical image segmentation,'' \emph{PloS one},
  vol.~17, no.~5, p. e0267976, 2022.

\bibitem{billot_synthseg_2021}
B.~Billot, D.~N. Greve, O.~Puonti, A.~Thielscher, K.~Van~Leemput, B.~Fischl,
  A.~V. Dalca, and J.~E. Iglesias, ``Synthseg: {Domain} {Randomisation} for
  {Segmentation} of {Brain} {MRI} {Scans} of any {Contrast} and {Resolution},''
  \emph{arXiv:2107.09559 [cs]}, 2021.

\bibitem{freesurfer}
B.~Fischl, ``Freesurfer,'' \emph{Neuroimage}, vol.~62, no.~2, pp. 774--781,
  2012.

\bibitem{fsl}
M.~Jenkinson, C.~F. Beckmann, T.~E. Behrens, M.~W. Woolrich, and S.~M. Smith,
  ``Fsl,'' \emph{Neuroimage}, vol.~62, no.~2, pp. 782--790, 2012.

\bibitem{brainsuite}
D.~W. Shattuck and R.~M. Leahy, ``Brainsuite: an automated cortical surface
  identification tool,'' \emph{Medical image analysis}, vol.~6, no.~2, pp.
  129--142, 2002.

\bibitem{volbrain}
J.~V. Manj{\'o}n and P.~Coup{\'e}, ``volbrain: an online mri brain volumetry
  system,'' \emph{Frontiers in neuroinformatics}, vol.~10, p.~30, 2016.

\bibitem{huo20193d}
Y.~Huo, Z.~Xu, Y.~Xiong, K.~Aboud, P.~Parvathaneni, S.~Bao, C.~Bermudez, S.~M.
  Resnick, L.~E. Cutting, and B.~A. Landman, ``3d whole brain segmentation
  using spatially localized atlas network tiles,'' \emph{NeuroImage}, vol. 194,
  pp. 105--119, 2019.

\bibitem{synthseg}
B.~Billot, D.~N. Greve, O.~Puonti, A.~Thielscher, K.~Van~Leemput, B.~Fischl,
  A.~V. Dalca, and J.~E. Iglesias, ``Synthseg: Domain randomisation for
  segmentation of brain mri scans of any contrast and resolution,'' \emph{arXiv
  preprint arXiv:2107.09559}, 2021.

\bibitem{fastsurfer}
L.~Henschel, S.~Conjeti, S.~Estrada, K.~Diers, B.~Fischl, and M.~Reuter,
  ``Fastsurfer-a fast and accurate deep learning based neuroimaging pipeline,''
  \emph{NeuroImage}, vol. 219, p. 117012, 2020.

\bibitem{quicknat}
A.~G. Roy, S.~Conjeti, N.~Navab, C.~Wachinger, A.~D.~N. Initiative
  \emph{et~al.}, ``Quicknat: A fully convolutional network for quick and
  accurate segmentation of neuroanatomy,'' \emph{NeuroImage}, vol. 186, pp.
  713--727, 2019.

\bibitem{deepnat}
C.~Wachinger, M.~Reuter, and T.~Klein, ``Deepnat: Deep convolutional neural
  network for segmenting neuroanatomy,'' \emph{NeuroImage}, vol. 170, pp.
  434--445, 2018.

\bibitem{hofmanninger2020automatic}
J.~Hofmanninger, F.~Prayer, J.~Pan, S.~R{\"o}hrich, H.~Prosch, and G.~Langs,
  ``Automatic lung segmentation in routine imaging is primarily a data
  diversity problem, not a methodology problem,'' \emph{European Radiology
  Experimental}, vol.~4, no.~1, pp. 1--13, 2020.

\bibitem{py06nimg}
P.~A. Yushkevich, J.~Piven, H.~Cody~Hazlett, R.~Gimpel~Smith, S.~Ho, J.~C. Gee,
  and G.~Gerig, ``User-guided {3D} active contour segmentation of anatomical
  structures: Significantly improved efficiency and reliability,''
  \emph{Neuroimage}, vol.~31, no.~3, pp. 1116--1128, 2006.

\bibitem{kikinis20133d}
R.~Kikinis, S.~D. Pieper, and K.~G. Vosburgh, ``3d slicer: a platform for
  subject-specific image analysis, visualization, and clinical support,'' in
  \emph{Intraoperative imaging and image-guided therapy}.\hskip 1em plus 0.5em
  minus 0.4em\relax Springer, 2013, pp. 277--289.

\bibitem{yeh2020shape}
F.-C. Yeh, ``Shape analysis of the human association pathways,''
  \emph{Neuroimage}, vol. 223, p. 117329, 2020.

\bibitem{brett_matthew_2023_7633628}
\BIBentryALTinterwordspacing
M.~Brett, C.~J. Markiewicz, M.~Hanke, M.-A. Côté, B.~Cipollini, P.~McCarthy,
  D.~Jarecka, C.~P. Cheng, Y.~O. Halchenko, M.~Cottaar, E.~Larson, S.~Ghosh,
  D.~Wassermann, S.~Gerhard, G.~R. Lee, H.-T. Wang, E.~Kastman, J.~Kaczmarzyk,
  R.~Guidotti, J.~Daniel, O.~Duek, A.~Rokem, C.~Madison,
  D.~Papadopoulos~Orfanos, A.~Sólon, B.~Moloney, F.~C. Morency, M.~Goncalves,
  Z.~Baratz, R.~Markello, C.~Riddell, C.~Burns, J.~Millman, A.~Gramfort,
  J.~Leppäkangas, J.~J. van~den Bosch, R.~D. Vincent, H.~Braun,
  K.~Subramaniam, A.~Van, K.~J. Gorgolewski, P.~R. Raamana, J.~Klug, B.~N.
  Nichols, E.~M. Baker, S.~Hayashi, B.~Pinsard, C.~Haselgrove, M.~Hymers,
  O.~Esteban, S.~Koudoro, F.~Pérez-García, J.~Dockès, N.~N. Oosterhof,
  B.~Amirbekian, I.~Nimmo-Smith, L.~Nguyen, S.~Reddigari, S.~St-Jean,
  E.~Panfilov, E.~Garyfallidis, G.~Varoquaux, J.~H. Legarreta, K.~S. Hahn,
  L.~Waller, O.~P. Hinds, B.~Fauber, J.~Roberts, J.-B. Poline, J.~Stutters,
  K.~Jordan, M.~Cieslak, M.~E. Moreno, T.~Hrnčiar, V.~Haenel, Y.~Schwartz,
  B.~C. Darwin, B.~Thirion, C.~Gauthier, I.~Solovey, I.~Gonzalez,
  J.~Palasubramaniam, J.~Lecher, K.~Leinweber, K.~Raktivan, M.~Calábková,
  P.~Fischer, P.~Gervais, S.~Gadde, T.~Ballinger, T.~Roos, V.~R. Reddam, and
  freec84, ``nipy/nibabel: 5.0.1,'' Feb. 2023. [Online]. Available:
  \url{https://doi.org/10.5281/zenodo.7633628}
\BIBentrySTDinterwordspacing

\bibitem{lowekamp2013design}
B.~C. Lowekamp, D.~T. Chen, L.~Ib{\'a}{\~n}ez, and D.~Blezek, ``The design of
  simpleitk,'' \emph{Frontiers in neuroinformatics}, vol.~7, p.~45, 2013.

\bibitem{diaz2021data}
O.~Diaz, K.~Kushibar, R.~Osuala, A.~Linardos, L.~Garrucho, L.~Igual, P.~Radeva,
  F.~Prior, P.~Gkontra, and K.~Lekadir, ``Data preparation for artificial
  intelligence in medical imaging: A comprehensive guide to open-access
  platforms and tools,'' \emph{Physica medica}, vol.~83, pp. 25--37, 2021.

\bibitem{despotovic2010brain}
I.~Despotovic, E.~Vansteenkiste, and W.~Philips, ``Brain volume segmentation in
  newborn infants using multi-modal mri with a low inter-slice resolution,'' in
  \emph{2010 Annual International Conference of the IEEE Engineering in
  Medicine and Biology}.\hskip 1em plus 0.5em minus 0.4em\relax IEEE, 2010, pp.
  5038--5041.

\bibitem{sled1998nonparametric}
J.~G. Sled, A.~P. Zijdenbos, and A.~C. Evans, ``A nonparametric method for
  automatic correction of intensity nonuniformity in mri data,'' \emph{IEEE
  transactions on medical imaging}, vol.~17, no.~1, pp. 87--97, 1998.

\bibitem{vnet}
F.~Milletari, N.~Navab, and S.-A. Ahmadi, ``V-net: Fully convolutional neural
  networks for volumetric medical image segmentation,'' in \emph{2016 Fourth
  International Conference on 3D Vision (3DV)}, 2016, pp. 565--571.

\bibitem{szegedy2017inception}
C.~Szegedy, S.~Ioffe, V.~Vanhoucke, and A.~A. Alemi, ``Inception-v4,
  inception-resnet and the impact of residual connections on learning,'' in
  \emph{Thirty-first AAAI conference on artificial intelligence}, 2017.

\bibitem{huang2017densely}
G.~Huang, Z.~Liu, L.~Van Der~Maaten, and K.~Q. Weinberger, ``Densely connected
  convolutional networks,'' in \emph{Proceedings of the IEEE conference on
  computer vision and pattern recognition}, 2017, pp. 4700--4708.

\bibitem{howard2017mobilenets}
A.~G. Howard, M.~Zhu, B.~Chen, D.~Kalenichenko, W.~Wang, T.~Weyand,
  M.~Andreetto, and H.~Adam, ``Mobilenets: Efficient convolutional neural
  networks for mobile vision applications,'' \emph{arXiv preprint
  arXiv:1704.04861}, 2017.

\bibitem{paszke2017automatic}
A.~Paszke, S.~Gross, S.~Chintala, G.~Chanan, E.~Yang, Z.~DeVito, Z.~Lin,
  A.~Desmaison, L.~Antiga, and A.~Lerer, ``Automatic differentiation in
  pytorch,'' in \emph{OpenReview}, 2017.

\bibitem{kingma2014adam}
D.~P. Kingma and J.~Ba, ``Adam: A method for stochastic optimization,''
  \emph{arXiv preprint arXiv:1412.6980}, 2014.

\bibitem{loshchilov2017decoupled}
I.~Loshchilov and F.~Hutter, ``Decoupled weight decay regularization,''
  \emph{arXiv preprint arXiv:1711.05101}, 2017.

\bibitem{taha2015metrics}
A.~A. Taha and A.~Hanbury, ``Metrics for evaluating {3D} medical image
  segmentation: analysis, selection, and tool,'' \emph{BMC medical imaging},
  vol.~15, no.~1, pp. 1--28, 2015.

\bibitem{razali2011power}
N.~M. Razali, Y.~B. Wah \emph{et~al.}, ``Power comparisons of shapiro-wilk,
  kolmogorov-smirnov, lilliefors and anderson-darling tests,'' \emph{Journal of
  statistical modeling and analytics}, vol.~2, no.~1, pp. 21--33, 2011.

\bibitem{gastwirth2009impact}
J.~L. Gastwirth, Y.~R. Gel, and W.~Miao, ``The impact of levene’s test of
  equality of variances on statistical theory and practice,'' \emph{Statistical
  Science}, vol.~24, no.~3, pp. 343--360, 2009.

\bibitem{nachar2008mann}
N.~Nachar \emph{et~al.}, ``The mann-whitney u: A test for assessing whether two
  independent samples come from the same distribution,'' \emph{Tutorials in
  quantitative Methods for Psychology}, vol.~4, no.~1, pp. 13--20, 2008.

\bibitem{hart2001mann}
A.~Hart, ``Mann-whitney test is not just a test of medians: differences in
  spread can be important,'' \emph{Bmj}, vol. 323, no. 7309, pp. 391--393,
  2001.

\bibitem{king2018statistical}
B.~M. King, P.~J. Rosopa, and E.~W. Minium, \emph{Statistical reasoning in the
  behavioral sciences}.\hskip 1em plus 0.5em minus 0.4em\relax John Wiley \&
  Sons, 2018.

\bibitem{antonelli2022medical}
M.~Antonelli, A.~Reinke, S.~Bakas, K.~Farahani, A.~Kopp-Schneider, B.~A.
  Landman, G.~Litjens, B.~Menze, O.~Ronneberger, R.~M. Summers \emph{et~al.},
  ``The medical segmentation decathlon,'' \emph{Nature Communications},
  vol.~13, no.~1, pp. 1--13, 2022.

\bibitem{ma_jun_2020_3757476}
\BIBentryALTinterwordspacing
M.~Jun, G.~Cheng, W.~Yixin, A.~Xingle, G.~Jiantao, Y.~Ziqi, Z.~Minqing, L.~Xin,
  D.~Xueyuan, C.~Shucheng, W.~Hao, M.~Sen, Y.~Xiaoyu, N.~Ziwei, L.~Chen, T.~Lu,
  Z.~Yuntao, Z.~Qiongjie, D.~Guoqiang, and H.~Jian, ``{COVID-19 CT Lung and
  Infection Segmentation Dataset},'' Apr. 2020. [Online]. Available:
  \url{https://doi.org/10.5281/zenodo.3757476}
\BIBentrySTDinterwordspacing

\end{thebibliography}

\end{document}